\documentclass[iop,revtex4]{emulateapj}
\usepackage{lineno}

\begin{document}

\renewcommand{\topfraction}{1.0}
\renewcommand{\bottomfraction}{1.0}
\renewcommand{\textfraction}{0.0}

\newcommand{\kms}{km~s$^{-1}$\,}
\newcommand{\msun}{$M_\odot$\,}

\title{Spectroscopic orbits of subsystems in  multiple
  stars. Vi.}

\author{Andrei Tokovinin}
\affil{Cerro Tololo Inter-American Observatory, Casilla 603, La Serena, Chile}
\email{atokovinin@ctio.noao.edu}

\begin{abstract}
Thirteen spectroscopic  orbits of late-type stars  are determined from
the high-resolution spectra taken with the CHIRON echelle spectrometer
at the 1.5-m  CTIO telescope.  Most (HIP 14194B,  40523A, 41171A, 51578A,
57572B, 59426A, 62852B, 66438A, 87813B, and 101472A) are inner subsystems
in  hierarchical multiple  stars with  three or  four  components. The
periods range from 2.2 to 1131 days. Masses of the components, orbital
inclinations,  and projected  rotation velocities  are  estimated, the
presence or absence of the lithium line is noted. In addition to those
systems, HIP~57021 is a simple 54-day twin binary, and HIP~111598 is a
compact triple-lined  system with  periods of 5.9  and 271  days. This
object is  likely old, but,  nevertheless, the secondary  component in
the  inner pair  does not  rotate synchronously  with the  orbit.  The
period-eccentricity diagram of  528 known inner low-mass spectroscopic
subsystems  (including  36 from  this  paper  series)  is given.   The
distribution  of the  inner periods  is smooth,  without  any details
around the tidal circularization period of $\sim$10\,d.
\end{abstract}

\maketitle

\section{Introduction}
\label{sec:intro}

Spectroscopic orbits  of  close binaries belonging  to hierarchical
stellar systems are determined  here, continuing the series of similar
papers   \citep{paper1,paper2,paper3,paper4,paper5}.     The   current
collection of  data on  stellar hierarchies, MSC \citep{MSC},  contains many
spectroscopic  binaries with  yet  unknown orbits.  Ongoing work  
slowly fills this gap.

\begin{deluxetable*}{c c rr   l cc rr r c }
\tabletypesize{\scriptsize}     
\tablecaption{Basic parameters of observed multiple systems
\label{tab:objects} }  
\tablewidth{0pt}                                   
\tablehead{                                                                     
\colhead{WDS} & 
\colhead{Comp.} &
\colhead{HIP} & 
\colhead{HD} & 
\colhead{Spectral} & 
\colhead{$V$} & 
\colhead{$V-K$} & 
\colhead{$\mu^*_\alpha$} & 
\colhead{$\mu_\delta$} & 
\colhead{RV} & 
\colhead{$\varpi$\tablenotemark{a}} \\
\colhead{(J2000)} & 
 & &   &  
\colhead{type} & 
\colhead{(mag)} &
\colhead{(mag)} &
\multicolumn{2}{c}{ (mas yr$^{-1}$)} &
\colhead{(km s$^{-1}$)} &
\colhead{(mas)} 
}
\startdata
03030$-$0205   & A  & 14194 & 18795 & F7V     & 6.33 & 1.18 & 104  &     6  & 35.3     & 17.40 \\   
               & B  & \ldots& \ldots& K3V?    & 9.97 & 2.66 & 108  &     2  & 36.12    & 17.61 \\   
08164$-$0314   & A  & 40523 & 69351 & F8      & 7.25 & 1.39 & $-$109&  12   & 28.53    & 12.82 \\
               & B  & \ldots& \ldots& \ldots  &10.39 &\ldots& $-$94 &$-$6   & 29.90    & 11.16 \\
08240$-$1548   & AB & 41171 & 70904 & F2/F3V  & 8.55 & 1.06 &  $-$26& $-$15 & $-$1.43  & 6.61 \\
10321$-$7005   & A  & 51578 & 91561 & F5V:    & 9.05 & 1.02 &  $-$9 &   2   & $-$3.89  & 7.85 \\
               & B  & \ldots& \ldots& \ldots  & 9.33 & 1.11 &  $-$9 &   3   & $-$4.2   & 7.85 \\
11414$-$4101   & A  & 57021 & 101614& G0V     & 6.87 & 1.46 & 166   &$-$124 & 16.37     & 28.80 \\
11480$-$6607   & A  & 57572 & 102579& K1V+... & 8.33 & 1.76 &  $-$268& 172  & 10.6:    & 26.95 \\
               & B  &  \ldots& \ldots&K6.5V   & 9.11 & 2.36 &  $-$279& 165  & 14.14    & 26.74 \\
12114$-$1647   & A  & 59426  & 105913 & K1    & 7.05 & 1.70 & $-$152& $-$47 & 2.37     & 29.67 \\
               & B  &  \ldots& \ldots& \ldots & 8.69 & \ldots&$-$152& $-$47 & 2.1      & 29.39 \\
12530+1502     & A  & 62852  & 111959& F5     & 7.93 & 0.46  &   $-$24&  29 & $-$8.1   & 6.17 \\
               & B  &  \ldots& \ldots& \ldots & 8.65 & 1.12  &  $-$24&  29  & $-$8.05  & 6.52 \\
13372$-$6142   & AB & 66438  & 118261 & F6V   & 5.63 & 1.20  & 147& $-$120  & 21.13    & 27.99\tablenotemark{b} \\
17563$-$1549   & A  & 87813  & 163336& A1V    & 5.89 & 0.19  & $-$1 & $-$73 &  $-$24.0 & 12.38 \\
               & B  & \ldots& \ldots& \ldots  & 8.89 & 1.18  &   1 & $-$65  &  $-$9.24 & 15.64 \\
20339$-$2710   & A  & 101472& 195719 & G8V    & 9.38 & 1.77  & 71  & $-$84  & $-$89.87 &  11.75 \\
               & B  &  \ldots& \ldots& \ldots & 11.97& 2.89  &  71  & $-$87 & $-$89.5  & 11.86 \\
22366$-$0034  & AB & 111598 & 214169 & G0III & 8.38 & 1.99 & 13     &$-$41 & 18.20 & 6.85 
\enddata
\tablenotetext{a}{Proper motions and parallaxes are taken
  from the {\it Gaia} DR2 \citep{Gaia}, where available.}
\tablenotetext{b}{{\it Hipparcos} parallax \citep{HIP2}.}
\end{deluxetable*}

The systems  studied here  are listed in  Table~\ref{tab:objects}. The
data  are  collected  from  Simbad  and {\it  Gaia}  DR2,  the  radial
velocities  (RVs) are mostly determined  here. The  first  column gives  the
WDS-style \citep{WDS} code based on the J2000 coordinates (most stars
are indeed present in the WDS). The HIP and HD identifiers, spectral
types, photometric and astrometric data refer either to the individual
stars or to the unresolved close subsystems. 

The structure of this paper is similar to the previous ones.  The data
and methods  are briefly recalled in  Section~\ref{sec:obs}, where the
 orbital elements are also  given. Then each system is discussed in
Section~\ref{sec:obj}.  The  paper  closes  with a  short  summary  in
Section~\ref{sec:sum}.

\section{Observations and data analysis}
\label{sec:obs}

\subsection{Spectroscopic observations}

The spectra used here were taken with the 1.5 m telescope sited at the
Cerro Tololo  Inter-American Observatory (CTIO) in  Chile and operated
by             the             SMARTS            Consortium.\footnote{
  \url{http://www.astro.yale.edu/smarts/}}  The   observing  time  (20
hours  per semester)  was allocated  through NOAO.   Observations were
made with  the CHIRON  optical echelle spectrograph  \citep{CHIRON} by
the telescope operators in service  mode.  The RVs are determined from
the cross-correlation function (CCF) of echelle orders with the binary
mask based on  the solar spectrum, as detailed  in \citep{paper1}. The
RVs derived  by this  method should  be on the  absolute scale  if the
wavelength  calibration  is accurate.   The  CHIRON  RVs were  checked
against standards and a small  offset of $+0.15$ km~s$^{-1}$ was found
in \cite{paper3};  it is not  applied to the  RVs given here.   The RV
errors depend  on the width and  contrast of the CCF  dip, presence of
other  dips,  signal  to  noise  ratio, and  other  factors.  The  rms
residuals from  the orbits can be  as low as 0.03  \kms, but typically
are between 0.1 and 0.2 \kms  for the systems studied here.  We assign
the RV  errors (hence  weights) to match  roughly the  residuals, with
larger errors assigned to blended or noisy dips. 

The CCF  contains two dips in  the case of  double-lined systems.  The
dip width is related to the projected rotation velocity $V \sin i$, while
its area depends on the spectral type, metallicity, and relative flux.
Table~\ref{tab:dip}  lists average parameters  of the  Gaussian curves
fitted to the CCF dips.   It gives the number of averaged measurements
$N$  (blended  CCFs  were  not  used),  the  dip  amplitude  $a$,  its
dispersion $\sigma$,  the product $a  \sigma$ proportional to  the dip
area (hence to the relative flux), and the projected rotation velocity
$V \sin i$,  estimated from $\sigma$ by the  approximate formula given
in \citep{paper1} and valid for $\sigma < 12$ \kms.  The last column
indicates  the presence  or absence  of the  lithium 6708\AA  ~line in
individual components.

\begin{deluxetable*}{l l c cccc c}    
\tabletypesize{\scriptsize}     
\tablecaption{CCF parameters
\label{tab:dip}          }
\tablewidth{0pt}                                   
\tablehead{                                                                     
\colhead{HIP} & 
\colhead{Comp.} & 
\colhead{$N$} & 
\colhead{$a$} & 
\colhead{$\sigma$} & 
\colhead{$a \sigma$} & 
\colhead{$V \sin i$ } & 
\colhead{Li}
\\
 &  &  & &
\colhead{(km~s$^{-1}$)} &
\colhead{(km~s$^{-1}$)} &
\colhead{(km~s$^{-1}$)} &
\colhead{  6708\AA}
}
\startdata
14194    & Ba & 5  & 0.258 &  4.588 &  1.182 &  5.5  & N \\
14194    & Bb & 5  & 0.201 &  4.656 &  0.934 &  5.7  & N \\
40523    & Aa & 5  & 0.158 &  4.162 &  0.659 &  4.3  & Y \\
40523    & Ab & 5  & 0.151 &  4.455 &  0.672 &  5.2  & Y \\
40523    & B  & 5  & 0.014 &  3.168 &  0.043 &  0.0  & \ldots  \\ 
41171    & Aa & 17 & 0.023 & 16.912 &  0.380 &  30: & Y? \\
41171    & Ab & 17 & 0.060 &  5.371 &  0.321 &  7.5  & Y? \\ 
51578    & Aa & 8  & 0.057 & 14.601 &  0.833 &  26: & N \\
51578    & B  & 2  & 0.035 & 24.050 &  0.853 &  43: & N \\
57021    & Aa & 12 & 0.167 &  3.474 &  0.579 &  1.3  & Y \\
57021    & Ab & 12 & 0.167 &  3.468 &  0.579 &  1.2  & Y \\
57572    & A  & 2  & 0.522 &  3.629 &  1.894 &  2.3  & N \\
57572    & Ba & 12 & 0.234 &  4.522 &  1.056 &  5.4  & N \\
57572    & Bb & 12 & 0.190 &  4.484 &  0.852 &  5.3  & N \\
59426    & Aa & 10 & 0.270 &  4.587 &  1.238 &  5.5  & Y \\
59426    & Ab & 10 & 0.161 &  3.742 &  0.604 &  2.8  & N \\
59426    & B  & 3  & 0.504 &  4.345 &  2.188 &  4.9  & N \\
62852    & Ba & 9 & 0.119  & 5.052  & 0.612  & 6.7   & Y \\
62852    & Bb & 9 & 0.103  & 5.624  & 0.577  & 8.1   & Y \\
66438    & Aa & 8 & 0.119  & 3.712  & 0.441  & 2.7   & Y \\
66438    & Ab & 8 & 0.073  & 3.455  & 0.252  & 1.1   & Y \\ 
66438    & B  & 8 & 0.061  & 7.687  & 0.471  & 12.4  & N? \\  
87813    & Ba & 1 & 0.206  & 7.183  & 1.476  & 11.4  & Y \\
87813    & Bb & 1 & 0.030  & 5.998  & 0.178  & 8.9   & Y? \\
101472   & Aa & 9 & 0.290  & 3.489  & 1.011  & 1.4   & N \\
101472   & Ab & 9 & 0.148  & 3.587  & 0.529  & 2.1   & N \\
111598   & Aa &24 & 0.049  & 15.608 & 0.771  &  27: & N \\
111598   & Ab &24 & 0.061  &  4.227 & 0.258  &  4.5  & N \\
111598   & B  &24 & 0.044  &  3.562 & 0.158  &  1.9  & N 
\enddata 
\end{deluxetable*}

\subsection{Orbit calculation}

As in the  previous papers of this series,  orbital elements and their
errors  were  determined  by   the  least-squares  fits  with  weights
inversely  proportional to  the  adopted errors.   The  IDL code  {\tt
  orbit}\footnote{Codebase:
  \url{http://www.ctio.noao.edu/\~{}atokovin/orbit/}                and
  \url{https://doi.org/10.5281/zenodo.61119} } was used \citep{orbit}.
It  can fit  spectroscopic, visual,  or  combined visual/spectroscopic
orbits. Formal  errors of orbital  elements are determined  from these
fits.    The   elements  of   spectroscopic   orbits   are  given   in
Table~\ref{tab:sborb}, in standard  notation. Its last column contains
the masses  $M \sin^3 i$ for double-lined  binaries.  For single-lined
systems, the  mass of  the primary star  (listed here with  colons) is
estimated from  its absolute $V$  magnitude, and the minimum  mass of
the secondary that corresponds  to the 90\degr ~inclination is derived
from the orbit.   Table~\ref{tab:rv}, published in ful electronically,
provides individual  RVs.  The {\it  Hipparcos} number of  the primary
star and  the system  identifier (components joined  by comma)  in the
first two columns  define the pair.  Then follow  the Julian date, the
RV, its  adopted error $\sigma$  (blended CCF dips are  assigned large
errors),  and the  residual to  the  orbit (O$-$C).   The last  column
specifies to which component this  RV refers ('a' for the primary, 'b'
for the  secondary, 'c' for the  tertiary of HIP~111598).   The RVs of
other   visual   components  are   provided,   for  completeness,   in
Table~\ref{tab:rvconst}.   It contains the  HIP number,  the component
letter, the Julian date, and the RV.

\begin{deluxetable*}{l l cccc ccc c c}    
\tabletypesize{\scriptsize}     
\tablecaption{Spectroscopic orbits
\label{tab:sborb}          }
\tablewidth{0pt}                                   
\tablehead{                                                                     
\colhead{HIP} & 
\colhead{System} & 
\colhead{$P$} & 
\colhead{$T$} & 
\colhead{$e$} & 
\colhead{$\omega_{\rm A}$ } & 
\colhead{$K_1$} & 
\colhead{$K_2$} & 
\colhead{$\gamma$} & 
\colhead{rms$_{1,2}$} &
\colhead{$M_{1,2} \sin^3 i$} 
\\
& & \colhead{(d)} &
\colhead{(JD +24,00,000)} & &
\colhead{(deg)} & 
\colhead{(km~s$^{-1}$)} &
\colhead{(km~s$^{-1}$)} &
\colhead{(km~s$^{-1}$)} &
\colhead{(km~s$^{-1}$)} &
\colhead{ (${\cal M}_\odot$) } 
}
\startdata
 14194 & Ba,Bb & 1136 & 58180.6    & 0.575 & 281.0 & 10.027 & 10.241 & 36.121 & 0.16 & 0.27 \\
      & & $\pm$5.8  & $\pm$3.7    & $\pm$0.023 & $\pm$2.1 & $\pm$0.356 & $\pm$0.363 & $\pm$0.062  & 0.18 & 0.27 \\
 40523 & Aa,Ab &28.9577     & 58427.130   & 0.524      & 228.0    & 48.994     & 49.088 & 28.531         & 0.12 & 0.87 \\
      &        &$\pm$0.0004 & $\pm$0.005  & $\pm$0.001 & $\pm$0.1 & $\pm$0.065 & $\pm$0.065 & $\pm$0.038 & 0.13 & 0.87 \\
 41171 &  Aa,Ab &25.4160  & 58449.990    & 0.533     & 308.1        &  46.739 & 48.325     & $-$1.438   & 1.67 & 0.70 \\
      &       &$\pm$0.0011 & $\pm$0.007  & $\pm$0.001 & $\pm$0.2 & $\pm$0.683 & $\pm$0.068 & $\pm$0.043 & 0.17 & 0.67 \\
 51578 &  Aa,Ab  &2.19189& 58300.390  & 0.078 & 237.1         & 40.087       & \ldots & $-$3.891  &  0.33 & 1.30: \\
      &  &$\pm$0.00001  & $\pm$0.012  &$\pm$0.004  &$\pm$2.0  & $\pm$0.132 & \ldots & $\pm$0.087 &  \ldots & $>$0.34 \\
 57021 &  Aa,Ab  & 53.6806 & 58498.043  & 0.462 & 55.9 & 33.930 & 33.906                    & 16.371     & 0.03  & 0.60  \\
      &       & $\pm$0.0029 & $\pm$0.0034 & $\pm$0.002 & $\pm$0.4 & $\pm$0.098 & $\pm$0.098 & $\pm$0.041 & 0.03  & 0.60 \\  
 57572 & Ba,Bb  &36.40406& 58315.672  & 0.142    & 122.5    & 35.463     & 36.337     & 14.143           & 0.06  & 0.68 \\
      &  &$\pm$0.0020 & $\pm$0.046  & $\pm$0.001 & $\pm$0.5 & $\pm$0.052 & $\pm$0.081 & $\pm$0.027 & 0.21 & 0.67 \\
 59426 & Aa,Ab   &211.585  & 58383.709  & 0.281  & 284.0    & 15.484     & 17.706     & 2.369      & 0.13 & 0.38 \\
      &  &$\pm$0.029  & $\pm$0.270  & $\pm$0.002 & $\pm$0.5 & $\pm$0.0039& $\pm$0.039 & $\pm$0.020 & 0.08 & 0.33 \\
 62852 &  Ba,Bb & 13.9288 & 58195.404  & 0.4324  & 207.6    & 66.903     & 67.747     & $-$8.050   & 0.10 & 1.30 \\
      &  &$\pm$0.0001 & $\pm$0.002  & $\pm$0.0005& $\pm$0.01 &$\pm$0.056 & $\pm$0.056 & $\pm$0.023 & 0.07 & 1.28 \\
 66438 & Aa,Ab   & 8.0714 & 58554.569 & 0        & 0        & 58.158     & 67.131     & 21.131     & 0.07 & 0.88   \\
      &  &$\pm$0.0002 & $\pm$0.0012 & fixed      & fixed    & $\pm$0.043 & $\pm$0.043 & $\pm$0.025 & 0.10 & 0.76 \\
 87813 & Ba,Bb & 728.96   & 53759.40   & 0.500    & 135.3     & 10.453     & 16.306     & $-$9.244   & 0.21 & 0.57 \\
      &  &$\pm$0.79   & $\pm$9.11    & $\pm$0.038 & $\pm$1.1  & $\pm$0.754 & $\pm$1.874 & $\pm$0.341 & 0.37 & 0.37 \\
 101472 & Aa,Ab  &354.88    & 58243.32  & 0.397      & 12.2 & 18.485      & 20.541      & $-$89.867  & 0.04 & 0.89 \\
      &          & $\pm$0.05 & $\pm$0.13 & $\pm$0.001 & $\pm$0.1  & $\pm$0.025  & $\pm$0.026  & $\pm$0.010 & 0.05 & 0.80 \\
 111598 & A,B      & 271.21  & 58560.65   & 0.230 & 306.3      & 14.224      & 29.777      & 18.199     & \ldots & 1.49 \\
      &  & $\pm$0.21    & $\pm$0.78    & $\pm$0.005& $\pm$0.9  & $\pm$0.132  & $\pm$0.239  & $\pm$0.037 & 0.26   & 0.71 \\
 111598 & Aa,Ab    & 5.86806 & 58383.4927 & 0     & 0    & 67.484      & 71.379      & \ldots     & 0.86 & 0.84 \\
      &  & $\pm$0.00003 & $\pm$0.0007  & fixed    & fixed   & $\pm$0.335  & $\pm$0.072  & \ldots  & 0.24 & 0.79 
\enddata 
\end{deluxetable*}



\begin{deluxetable}{r l c rrr c }    
\tabletypesize{\scriptsize}     
\tablecaption{Radial velocities and residuals (fragment)
\label{tab:rv}          }
\tablewidth{0pt}                                   
\tablehead{                                                                     
\colhead{HIP} & 
\colhead{System} & 
\colhead{Date} & 
\colhead{RV} & 
\colhead{$\sigma$} & 
\colhead{(O$-$C) } &
\colhead{Comp.}  \\
 & & 
\colhead{(JD +24,00,000)} &
\multicolumn{3}{c}{(km s$^{-1}$)} 
}
\startdata
 14194 &Ba,Bb  &  56922.7662 &   27.13 &    0.20 &   $-$0.10 &  a \\
 14194 &Ba,Bb  &  56922.7662 &   45.13 &    0.20 &   $-$0.07 &  b \\
 14194 &Ba,Bb  &  56937.7094 &   27.04 &    0.20 &   $-$0.15 &  a \\
 14194 &Ba,Bb  &  56937.7094 &   45.30 &    0.20 &      0.06 &  b 
\enddata 
\end{deluxetable}

\begin{deluxetable}{r l r r }    
\tabletypesize{\scriptsize}     
\tablecaption{Radial velocities of other components
\label{tab:rvconst}          }
\tablewidth{0pt}                                   
\tablehead{                                                                     
\colhead{HIP} & 
\colhead{Comp.} & 
\colhead{Date} & 
\colhead{RV}   \\ 
 & & 
\colhead{(JD +24,00,000)} &
\colhead {(km s$^{-1}$)}  
}
\startdata
40523 & B &   58228.4952 & 29.936 \\ 
40523 & B &   58232.4856 & 30.119 \\
40523 & B &   58454.7632 & 30.203 \\ 
40523 & B &   58456.7519 & 29.369 \\
40523 & B &   58484.7321 & 29.885 \\
40523 & B &   58513.6322 & 30.078 \\
40523 & B &   58526.6080 & 29.707 \\
51578 & B &   58194.6145 & $-$4.2:  \\
51578 & B &   58195.6229 & $-$3.7: \\
57572 & A &   58193.7407 & 10.658 \\
57572 & A &   58195.6377 & 10.456 \\
62852 & A &   58194.7057 & $-$5.4: \\
62852 & A &   58195.7036 & $-$6.4: \\
66438 & B &   58559.6566 & 21.199 \\
66438 & B &   58565.7035 & 21.227 \\
66438 & B &   58578.7043 & 21.213 \\
66438 & B &   58581.8054 & 21.233 \\
66438 & B &   58583.7687 & 21.184 \\
66438 & B &   58594.7517 & 21.169 \\
66438 & B &   58668.6582 & 20.987 \\
66438 & B &   58699.5444 & 21.214 
\enddata 
\end{deluxetable}

\subsection{Gaia DR2 astrometry}

Rich  data  contained in  the  second  {\it  Gaia} data  release,  DR2
\citep{Gaia},  contribute to the  study of  multiple stars  in various
ways. Here I  use the potential of detecting  astrometric subsystem by
the difference of  the short-term proper motion (PM)  measured by {\it
  Gaia}  with  the long-term  PM  $\mu_{\rm  mean}$  deduced from  the
comparison  between  the  {\it  Gaia} and  {\it  Hipparcos}  positions
\citep{Brandt2018}. The long-term PM is very accurate. A statistically
significant difference $\Delta  \mu_{\rm DR2-mean}$ indicates presence
of an  astrometric subsystems with  period from a  few years to  a few
decades.

\section{Individual objects}
\label{sec:obj}

For each observed system, the corresponding Figure shows a typical CCF
(the Julian  date and  individual components are  marked on  the plot)
together with the RV curve  representing the orbit.  In the RV curves,
squares denote  the primary component, triangles  denote the secondary
component, while the  full and dashed lines plot  the orbit. Masses of
stars are estimated from  absolute magnitudes, orbital periods of wide
pairs  from  their   projected  separations  \citep[see][]{MSC}.  Some
results  of  speckle   interferometry  at  the  Southern  Astrophysical
Research Telescope (SOAR) are used \citep{SOAR}.

\subsection{HIP 14194 (Triple)}

\begin{figure} 
\epsscale{1.0}
\plotone{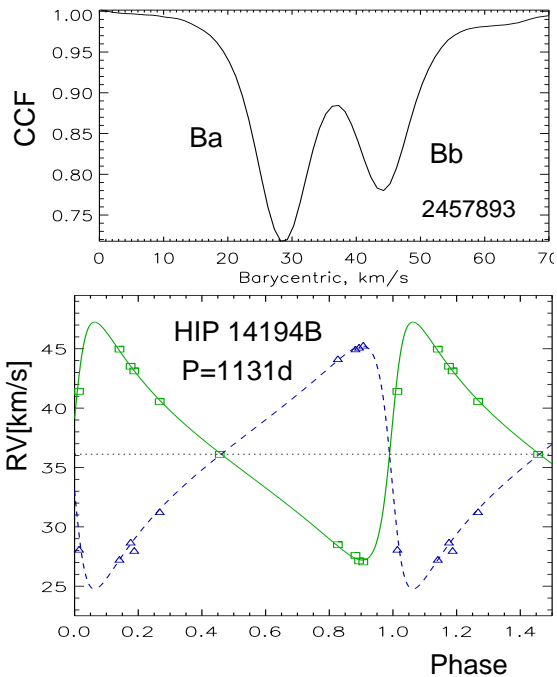}
\caption{CCF (top)  and RV curve (bottom)  of HIP 14194  Ba,Bb. In this
  and the following Figures, the upper panel shows a typical CCF, with
  the Julian date of the observation indicated.  The lower  panel  presents the
  phased RV  curve where the full  lines and squares  denote the orbit
  and  RVs of  the primary  component, the  dashed line  and triangles
  refer to the secondary component.
\label{fig:14194}
}
\end{figure}

The outer  visual pair  STF~341 was discovered  by W.~Struve  in 1831.
Its  estimated period  is $\sim$8  kyr;  the system  turned by  9\degr
~during  188 yrs  since  its  discovery at  a  constant separation  of
8\farcs7.  The {\it Gaia} parallaxes  and PMs of both components A and
B match perfectly; the star A is located on the main sequence, while B
is above it.  There is no significant $\Delta \mu_{\rm DR2-mean}$. The
star A was targeted by exo-planet RV surveys.

The  double-lined  nature  of   B  has  been  discovered  with  CHIRON
\citep{survey}.  Here its spectroscopic orbit  with a period of 3.1 yr
is  determined  (Figure~\ref{fig:14194}).   The  period  and  parallax
correspond to  the semimajor  axis of 43\,mas,  making the  Ba,Bb pair
accessible to speckle interferometry.  Indeed, it has been resolved at
SOAR  in 2018  at  similar  separation and  shows an orbital motion  in
agreement with  the short  period.  The speckle  measurements combined
with  the RVs  lead to  the preliminary  visual  elements $a=41$\,mas,
$\Omega =  162$\degr, $i=139$\degr. The masses  of Ba and  Bb are 0.76
and 0.74 \msun and their sum,  1.49 \msun, is similar to the estimated
mass of the component A,  1.22 \msun. The orbital inclination of Ba,Bb
evaluated  from the  spectroscopic masses  is  45\degr  ~or 135\degr;  the
latter agrees with the provisional visual elements.

\subsection{HIP 40523 (Triple)}

\begin{figure}
\epsscale{1.0}
\plotone{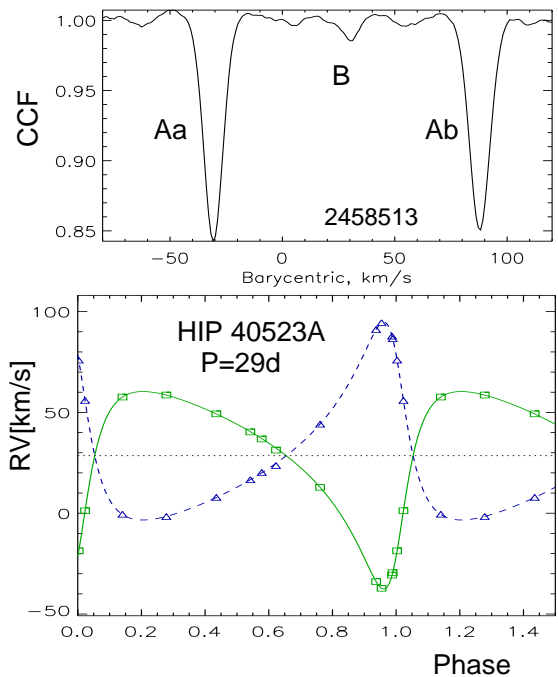}
\caption{CCF (top)  and RV curve (bottom) of HIP 40523 Aa,Ab.
\label{fig:40523}
}
\end{figure}

The  outer 1\farcs4 visual  pair A,B  (HWE~21, ADS~6707),  known since
1879, has an  estimated period of $\sim$600 yr and  has turned only by
21\degr ~since  its discovery.  The companion  B is fainter  than A by
3.5 mag and its RV is barely  measurable from the weak dip in the CCF;
the mean RV(B) and the rms  scatter are 29.90 and 0.29 \kms, respectively.
On the  other hand, the dips  belonging to the inner  binary Aa,Ab are
strong  and   equal.   Its  double-lined  nature   was  discovered  by
\citet{N04}.  The orbit of this  pair determined here has $P= 28.96$ d
and $e=  0.524$  (Figure~\ref{fig:40523}).  The  mass ratio $q=0.998$
makes the inner pair a perfect twin. The estimated masses of Aa and Ab
are 1.29  \msun, hence the  inclination is $i_{\rm Aa,Ab}  = 62\degr$.
Considering the detectable lithium  line and the kinematics typical of
a young  disk population, $(U,V,W)  = (-43.0, -9.9 -23.1)$  \kms, this
system could be relatively young.

\subsection{HIP 41171 (Quadruple)}

\begin{figure}
\epsscale{1.0}
\plotone{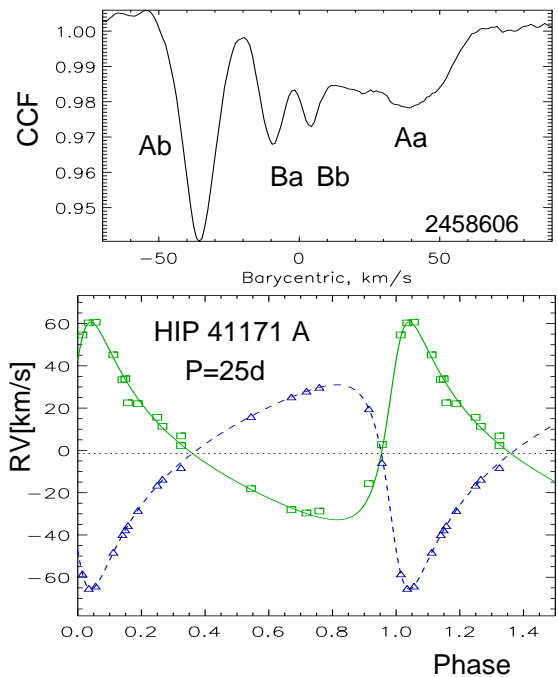}
\caption{CCF (top)  and RV curve (bottom) of HIP 41171 Aa,Ab.
\label{fig:41171}
}
\end{figure}

The outer  visual binary  A,B discovered by  R.~A.  Rossiter  in 1940
(RST~4396) has  a separation of  1\arcsec ~and an estimated  period of
$\sim$1 kyr.   The magnitude difference  between A and B  is moderate,
about 2 mag.  All  available photometry and astrometry, including {\it
  Gaia}, refers to the unresolved pair AB. Understandably, there is no
significant $\Delta \mu$ because the  outer period is too long and the
inner period too short for detecting an astrometric acceleration.

\citet{N04}   discovered  that  the   pair  contains   a  double-lined
spectroscopic subsystem  and the  stars are slightly  metal deficient,
[Fe/H]=$-$0.2  dex. Our  first observations  revealed CCFs  with three
dips belonging to  Aa, Ab, and B.   The dip of Aa is  wide and shallow
owing to its fast axial rotation, while the dips of Ab and B are much
sharper. However, further monitoring  revealed that the component B is
also a double-lined binary: its dip separated in two components Ba and
Bb, with  a slow change of RVs  (Figure~\ref{fig:41171}).  The orbital
period of Ba,Bb exceeds one year, and  observations will continue  in
the next season, selecting dates when  the dips of Aa and Ab do not
overlap with the  dips of B.  I do  not report here the RVs  of B and
the parameters of  its CCF dips. The area of the  blended dip of Ba+Bb
is used to derive relative fluxes.

The RV  of A is measured poorly  and its residuals from  the orbit are
large, so the  elements of Aa,Ab are based mostly  on the accurate RVs
of Ab.   Although the dip areas of  Aa, Ab, and B  are comparable, the
actual  relative  fluxes differ  more  substantially because  spectral
lines are  stronger in cool stars. Adopting  effective temperatures of
6500, 6400,  and 5900 K  for Aa, Ab,  and B, respectively,  I estimate
their individual  $V$ magnitudes  as 9.41, 9.65,  and 10.39  mag.  The
resulting $\Delta V_{\rm AB} = 1.61$ mag is a little less than $\Delta
Hp_{\rm A,B} = 1.65$ mag  from {\it Hipparcos} and $\Delta I_{\rm A,B}
= 1.8$ mag from  SOAR speckle interferometry.  The absolute magnitudes
of  Aa  and Ab  match  main-sequence masses  of  1.28  and 1.23  \msun
(spectral  types F5V  and F6V),  in agreement  with the  measured mass
ratio $q_{\rm  Aa,Ab} = 0.97$.   The resulting orbital  inclination of
the Aa,Ab  pair is $i_{\rm  Aa,Ab} \approx 54^\circ$.  These estimates
are only preliminary.

Considering the likely  presence of a weak lithium lines  in the spectra
of Aa  and Ab and the  measurable rotation of these  stars, the system
can be relatively  young.  The faster rotation of  Aa can be explained
by its earlier spectral type,  hence a shallower convective zone and a
longer braking time compared to Ab.

\subsection{HIP 51578 (Triple)}

\begin{figure}
\epsscale{1.0}
\plotone{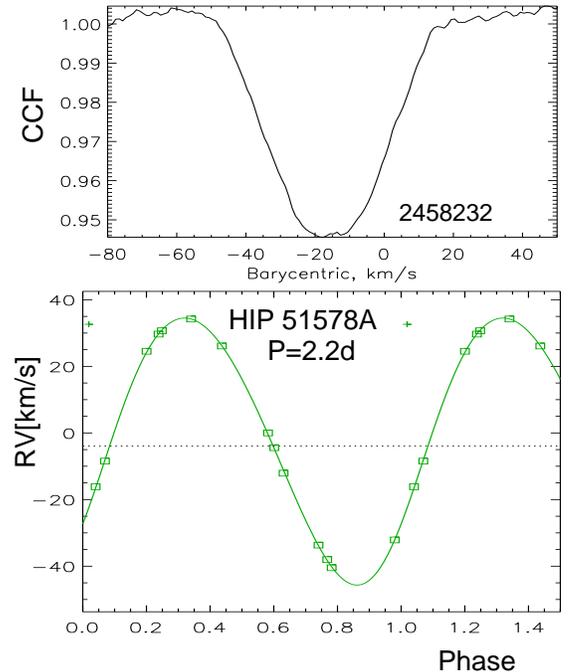}
\caption{CCF (top) and RV curve (bottom) of HIP 51578 Aa,Ab. The cross shows
  RV from \citet{Desidera2006}, not used to fit the elements. 
\label{fig:51578}
}
\end{figure}

The outer  7\farcs8 binary HJ~4335, discovered by  J.~Hershel in 1834,
has an estimated period of $\sim$20  kyr. The stars A and B are almost
equal,  $\Delta  V   =  0.29$  mag,  and  are   located  on  the  main
sequence. The component A has no significant $\Delta \mu$. The RV of B
measured by  CHIRON ($-$4.2  and $-$3.7 \kms)  and {\it  Gaia} ($-$2.9
\kms) is close to the center-of-mass RV of A, $-3.4$ \kms. Owing to
the fast rotation of B, its RVs are not accurate. 

The RV of A is variable, as established by \citet{N04}.  The period is
short, $P=2.192$  d, but  the orbit is  definitely non-circular,  $e =
0.078  \pm  0.004$  (Figure~\ref{fig:51578}).   One   RV  measured  by
\citet{Desidera2006}  can be  fitted  by a  slight  adjustment of  the
period, but this increases  the residuals substantially, from 0.33 \kms
to  0.75 \kms.  Ongoing  tidal circularization  of an  eccentric orbit
with a slight  decrease in period offers potential  explanation for the
deviant RV.

\begin{figure}
\epsscale{1.0}
\plotone{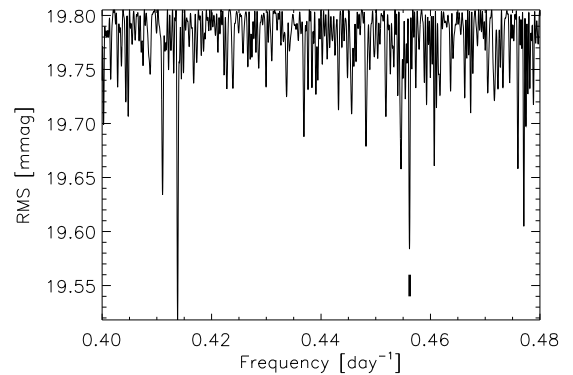}
\caption{Periodogram of HIP~51578 AB. The orbital frequency of 0.4562
  day$^{-1}$ is marked by the thick line. The strongest dip is at
  0.4137  day$^{-1}$. 
\label{fig:periodogram}
}
\end{figure}

I  used the  ASAS-3 photometry  in the  $V$ band
\citep{ASAS}\footnote{ \url{http://www.astrouw.edu.pl/asas/?page=catalogues}}
to  search for
photometric signal with the orbital period. The components A and B are
measured  jointly.   The  plot in  Figure~\ref{fig:periodogram}  shows
the residual  rms variation  after subtracting  the best-fitting  sine and
cosine terms at each trial  frequency.  The strongest feature is found
at the frequency of 0.4137  day$^{-1}$ (period 2.417 d); it implies an
rms variation of only 3.22 mmag  and may  correspond to the rotation
of the  star B.  There are also  dips at the orbital  frequency and at
its  alias, possibly caused  by the  starspots synchronized  with the
orbit of Aa,Ab.  The ASAS photometry proves the absence of eclipses in
the subsystem. 

The minimum mass of the spectroscopic secondary Ab is 0.34 \msun.  The
projected rotation $V \sin i$ for Aa and B estimated crudely from the width of
the CCF dips,  26 and 43 \kms respectively,  agrees  with 25.1
and  39.4  \kms  measured  by  \citet{Desidera2006}.  The  synchronous
rotation speed of  a star of 1.3 $R_\odot$ radius  is 29.7 \kms, hence
the orbital inclination of Aa,Ab is $\sim$60\degr ~and the actual mass
of Ab is about 0.4 \msun.

\subsection{HIP 57021 (Binary)}

\begin{figure}
\epsscale{1.0}
\plotone{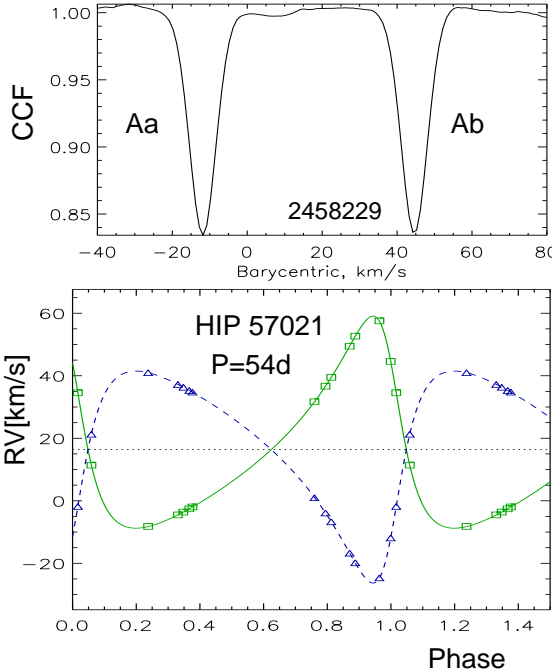}
\caption{CCF (top)  and RV curve (bottom) of HIP 57021.
\label{fig:57021}
}
\end{figure}

This is just  a nearby (35\,pc) spectroscopic binary  belonging to the
67-pc  sample of  solar-type stars  \citep{FG67b}.  Double  lines were
detected by \citet{N04}, but the orbital period remained unknown.  The
components  are  practically  indistinguishable:  the  mass  ratio  is
1.0007$\pm$0.0042 and the  dips are equal. The period  of this perfect
twin is  53.7 days, well  above the tidal circularization  limit. Note
the  small residuals  of  0.03  \kms. The  orbital  inclination is  $i
\approx 60\degr$.  The projected axial  rotation is very slow, $V \sin
i \approx 1.3$ \kms. Nevertheless, lithium is not yet destroyed in the
atmosphere of these  G0V dwarfs and they are  featured in the catalogs
of chromospherically active stars \citep[e.g.][]{Boro2018}.   The kinematics does not distinguish this star from the disk population, $(U,V,W) = (31.6, 6.6, -11.0)$ \kms. \citet{Fuhrman2019} measured the metallicity [Fe/H]=$-0.41$ dex. \citet{Brandt2018} did not detect any significant acceleration which, together with very small RV residuals, indicates the absence of additional components with periods less than $\sim$30\,yr.

\subsection{HIP 57572 (Quadruple)}

\begin{figure}
\epsscale{1.0}
\plotone{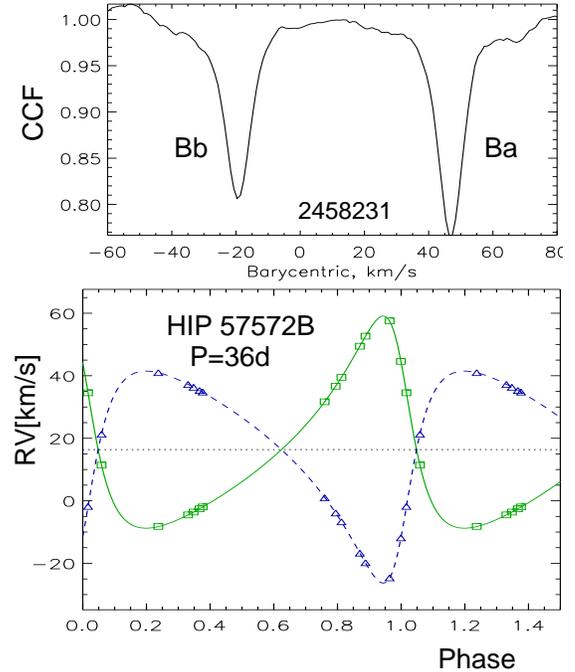}
\caption{CCF (top)  and RV curve (bottom) of HIP 57572 Ba,Bb.
\label{fig:57572}
}
\end{figure}

This is a low-mass nearby (37\,pc) stellar system with a fast PM.  The
outer 8\farcs2 binary  GLI~168 was first resolved in  1852. Its period
is $\sim$3.5 kyr.  The secondary star  B, 1.4 mag fainter than A, is a
double-lined  spectroscopic binary  \citep{N04}. Here  its  orbit with
$P=36$ d  is determined (Figure~\ref{fig:57572}). This  is yet another
twin  with $q=0.976$. The  masses of  Ba and  Bb estimated  from their
absolute magnitudes, $\sim$0.7 \msun,  are close to the spectroscopic
masses $M \sin^3 i$, hence the orbit of Ba,Bb is seen edge-on.

\citet{N04} measured  the mean RV(A)  of 8.8 \kms  and found it  to be
variable by 1.9 \kms over a 2-yr period.  The variability is confirmed
by the different RVs measured  with CHIRON, 10.5 \kms, and {\it Gaia},
12.8  \kms (the latter  with a  large error  of 1.8  \kms), and  by the
difference with the center-of-mass RV  of B, 14.14 \kms.  Therefore, A
likely  contains a  low-mass satellite  with yet  unknown  period.  An
indirect evidence of  the existence of a subsystem  is the large error
of  the DR2  parallax  of A,  0.3\,mas,  while the  parallax  of B  is
measured by  {\it Gaia} with an error  of 0.03 mas.  There  is a small
acceleration  $\Delta  \mu_{\rm  DR2-mean}  \approx  2$  mas~yr$^{-1}$.   The
observed  motion  of the  pair  A,B  may show  a  wave  caused by  the
subsystem Aa,Ab.

\subsection{HIP 59426 (Triple)}

\begin{figure}
\epsscale{1.0}
\plotone{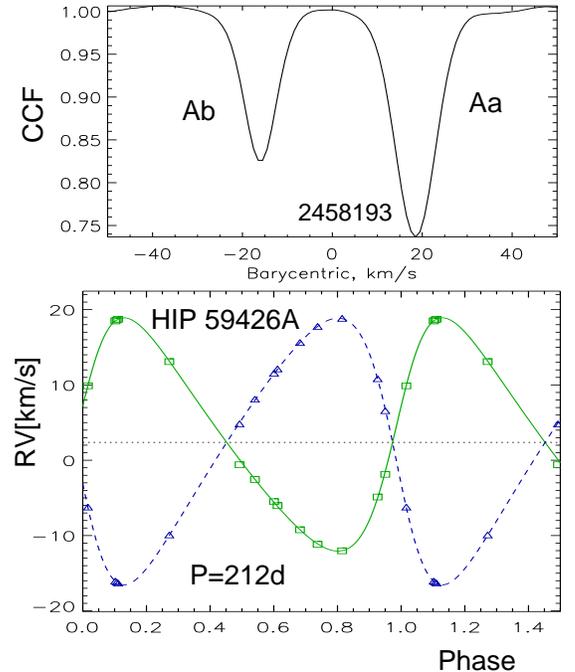}
\caption{CCF (top)  and RV curve (bottom) of HIP 59426 Aa,Ab.
\label{fig:59426}
}
\end{figure}

Like  the previous  object, this  system is  a nearby  (34  pc) triple
K-dwarf with a fast PM. The  outer pair A,B was first measured in 1877
at 6\farcs9 (S~634,  ADS~8444) and has closed down  to 4\farcs7 since.
Its estimated  period is 1.2 kyr.   Double lines in the  spectrum of A
were noted  by \citet{N04}.  Monitoring with CHIRON,  started in 2015,
revealed  a  relatively long  period,  $P=212$  d,  and a  mass  ratio
$q=0.874$  (Figure~\ref{fig:59426}). The  estimated semimajor  axis of
the inner orbit is 27\,mas.   The subsystem Aa,Ab was resolved in 2018
and  2019 by speckle  interferometry at  SOAR at  similar separations,
just under the formal diffraction limit.  The astrometric acceleration
$ \Delta  \mu_{\rm DR2-mean} = (-5.90, -0.2)$  mas~yr$^{-1}$ possibly results
from the orbital motion of Aa,Ab. On the other hand, the PM difference
between A and B matches the observed relative motion of the outer pair
with the average  velocity of 21 mas~yr$^{-1}$ (or 3.4 \kms)  in the plane of
the sky.

Interestingly, the  lithium 6708\AA ~line is securely  detected in the
spectra of  Aa and is equally  securely  missing  in both Ab and  B. As
lithium destruction depends on both  age and mass, this datum can help
in establishing the age of  this system. Its spatial velocity $(U,V,W)
= (-17.4, -17.4, -7.3)$ \kms is  close to the motion of young stars in
the solar vicinity, such as the  TW~Hya group. However, the orbital motion of A
in the  wide pair  has not been  subtracted from the  calculated space
velocity. 


\subsection{HIP 62852 (Quadruple)}

\begin{figure}
\epsscale{1.0}
\plotone{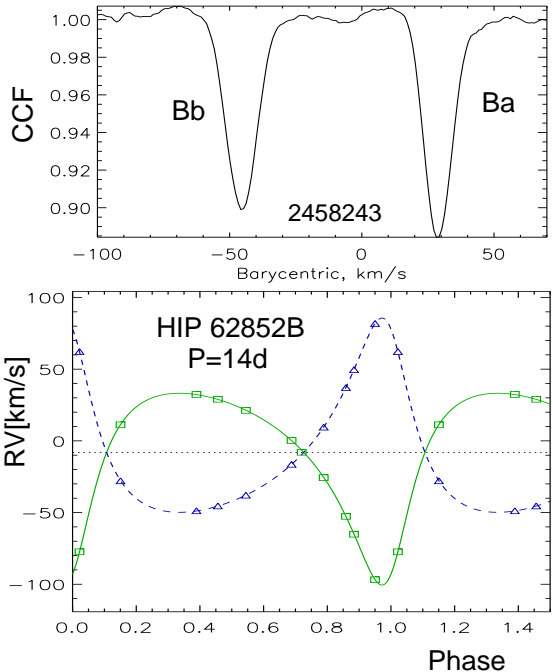}
\caption{CCF (top)  and RV curve (bottom) of HIP 62852 Ba,Bb.
\label{fig:62852}
}
\end{figure}

This is  2+2 quadruple system where  two close pairs  revolve around a
common center on a wide orbit  with an estimated period of 13 kyr. The
outer 5\farcs7 visual pair STF~1686  (ADS 8693) has been discovered by
W.~Struve  in 1823.  However,  its first  measurement is  misleadingly
inaccurate  and leaves  an impression  that the  pair moves  too fast.
{\it Gaia}  establishes the physical  nature of the outer  pair beyond
any doubt. The  WDS lists another faint companion  D at 1\farcs88 from
A, but it is not detected by {\it Gaia}, so I assume that D is just a
background star.

The  component  A  was  resolved  by  speckle  interferometry  into  a
0\farcs12 pair  Aa,Ab (YSC~215) with  an estimated period  of $\sim$60
yr.   The  visual  secondary,  B,  is a double-lined  pair  reported  by
\citet{N04}.   Here  its  spectroscopic  orbit  with  $P=13.9$  d  and
$q=0.987$  is   determined  (Figure~\ref{fig:62852}).   Its  estimated
inclination  is 80\degr. The  spectra of  Ba and  Bb have  the lithium
line,   and   these   stars    rotate   slightly   faster   than   the
pseudo-synchronous speed.


\subsection{HIP 66438 (Triple)}

\begin{figure}
\epsscale{1.0}
\plotone{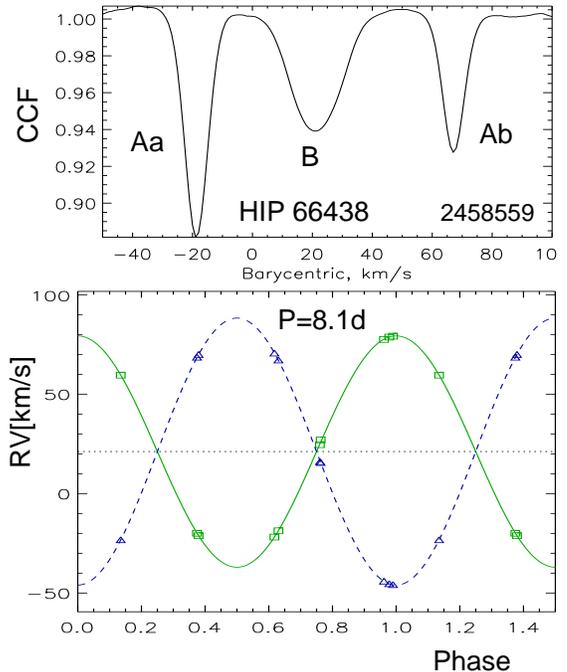}
\caption{CCF (top)  and RV curve (bottom) of HIP 66438 Aa,Ab.
\label{fig:66438}
}
\end{figure}

This is  a bright (HR~5113)  and nearby (36\,pc) tight  triple system.
The outer  visual pair A,B (I~365AB) was  discovered by R.~T.~A.~Innes
in  1900.  Its  orbit with  a period  of 35.0\,yr  is of  high quality
(grade 2).  The distant component C listed in the WDS is optical.   The
spectroscopic      subsystem      has      been     discovered      by
\citet{Evans1968,Fekel1981,Fuhrman2019}  by  the  presence  of  triple
lines in the spectrum.  The latter authors estimated 
the masses of the components as  1.21, 1.19, and 1.09 \msun ~and noted
the  presence of  the lithium  lines  in Aa  and Ab.   I confirm  this
finding; on the other hand, the  lithium line in the spectrum of B can
be only guessed.  The presence of lithium and  a relatively fast axial
rotation of  B (12.4 \kms) suggest  youth and are in  tension with the
age of  2.3\,Gyr estimated by \citet{Fuhrman2019}.   Relative areas of
the CCF  dips correspond  to $\Delta  m_{\rm A,B} =  0.42$ mag  in the
outer pair, while WDS gives $\Delta m_{\rm A,B} = 0.60$ mag, confirmed
by  the  speckle  interferometry  at  SOAR.   The  spectroscopic  pair
certainly belongs to the brighter visual component A.

Our circular orbit with a  period of 8.07\,d corresponds to the masses
$M \sin^3  i$ of 0.88 and 0.76  \msun. An attempt to  fit an eccentric
orbit does not improve the residuals and the resulting eccentricity of
0.003 is  not significant, hence  the circular orbit is  imposed.  The
masses  quoted above (which  match the  outer mass  sum of  3.39 \msun
derived from the visual orbit  and the {\it Hipparcos} parallax) yield
the inner inclination of 64\degr ~or 117\degr.  The inclination of the
outer orbit  is 117\degr, hinting at  possible co-planarity.  However,
the  outer  orbit  has  a  high eccentricity  $e=0.78$.  


The  outer  pair is  presently  near  the  apastron of  its  eccentric
orbit. The mean RV(B) is 21.18 \kms (rms scatter 0.08 \kms), very close
to the  center of mass  RV(A), 21.13 \kms.  Further  monitoring should
reveal the RV variation of both A and B caused by the eccentric 35-yr
outer orbit. The next periastron passage will be in 2037. 

\subsection{HIP 87813 (Quadruple)}

\begin{figure}
\epsscale{1.0}
\plotone{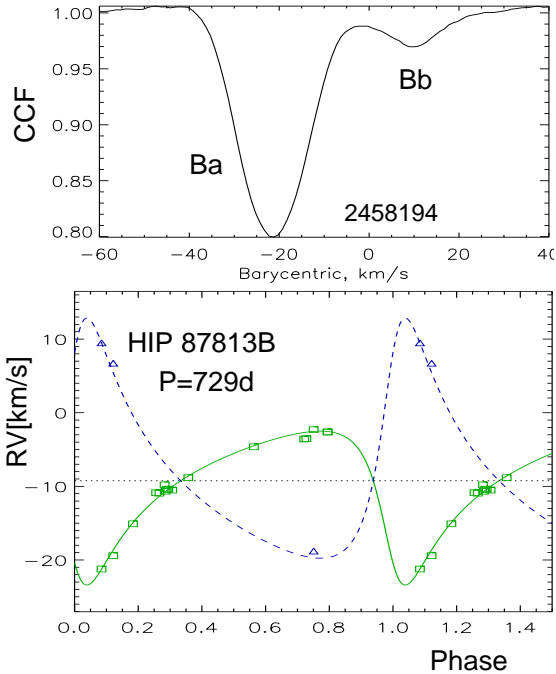}
\caption{CCF (top)  and RV curve (bottom) of HIP 87813 Ba,Bb.
\label{fig:87813}
}
\end{figure}

This  is a 2+2  quadruple system.   The outer  pair A,B  (HJ~2814, ADS
10891)  at 20\farcs6  separation is  known since  1830;  its estimated
period  is $\sim$30  kyr.   The  main component  A  (HR~6681), of  A1V
spectral  type,  has  variable   RV  according  to  \citet{N85}.   The
short-term PMs  of A measured by  both {\it Gaia}  and {\it Hipparcos}
differ significantly  from the mean  PM: $\Delta \mu_{\rm  DR2-mean} =
(-3.0, -1.5)$  mas~yr$^{-1}$ and $\Delta \mu_{\rm HIP2-mean}  = (-5.3, +6.0)$
mas~yr$^{-1}$.  The period of the subsystem Aa,Ab is likely a few years.

The RV of the  secondary component B (BD$-$15$^\circ$4723, ADS~10891B)
was  found to  be variable  by  \citet{TS02}; a  tentative orbit  with
$P=761$ d based  on those RVs was featured in the  old versions of the
MSC.   Now, using  additional  RVs  from CHIRON  and  Du Pont  echelle
\citep{LCO},  a   definitive  orbit  with  $P=729$   d  is  determined
(Figure~\ref{fig:87813}).   Moreover,  a  weak  dip  produced  by  the
secondary  component Bb  is  detected, allowing  to  measure the  mass
ratio, $q_{\rm Ba,Bb} = 0.82$.   The color and absolute magnitude of B
correspond to a  G0V star with a mass of  1.1 \msun.   Estimated masses
lead to the inclination of  $i_{\rm Ba,Bb} \approx 53\degr$.  The lithium line
is present in the spectrum of Ba.

\subsection{HIP 101472 (Triple)}

\begin{figure}
\epsscale{1.0}
\plotone{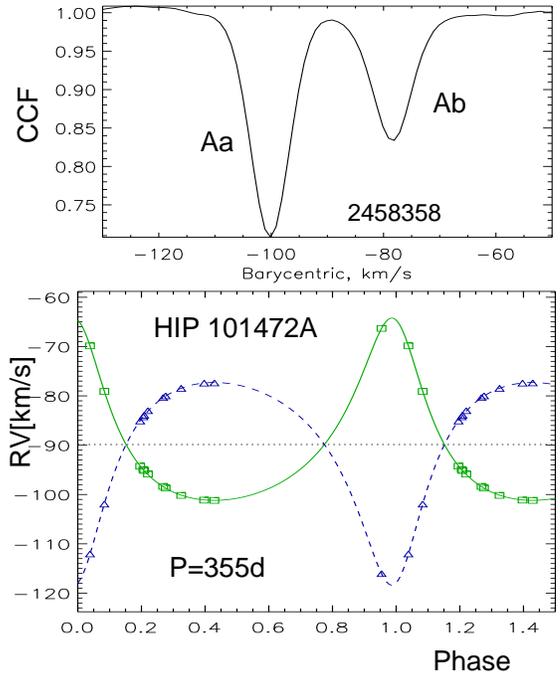}
\caption{CCF (top)  and RV curve (bottom) of HIP 101472 Aa,Ab.
\label{fig:101472}
}
\end{figure}

In this triple system, the outer  orbit has a long period of $\sim$200
kyr and  a wide separation of  52\arcsec. The components A  and B have
common parallax, PM,  and RV and are located on  the main sequence. No
astrometric acceleration of A can be detected.

Double  lines  in  the  component  A were  noted  by  \citet{N04}  who
estimated the mass  ratio of 0.92.  The star  was observed with CHIRON
since 2015,  showing double CCF  dips with slow variation.   The orbit
has  a  period   close  to  one  year  and   $q_{\rm  Aa,Ab}  =  0.92$
(Figure~\ref{fig:101472}).  Its semimajor axis  is 14 mas, so the pair
can  be resolved  by  speckle interferometry  at  8-m telescopes.  The
estimated inclination  is $i_{\rm Aa,Ab} = 77\degr$.   Presence of a
1-yr subsystem, unaccounted for by  the {\it Gaia} DR2 astrometry, can
bias the parallax, but this has not happened in this case. The spatial
velocity $(U,V,W) = (-84.0,  -51.2, 18.6)$ \kms and [Fe/H]=$-$0.23 dex
indicate  that the  system  may belong  to  the thick  disk (note  the
relatively large systemic RV of $-89.8$ \kms); slow axial rotation and
the absence of lithium support this view.

\subsection{HIP 111598 (Triple)}
\label{sec:111598}

\begin{figure*}
\epsscale{1.0}
\plotone{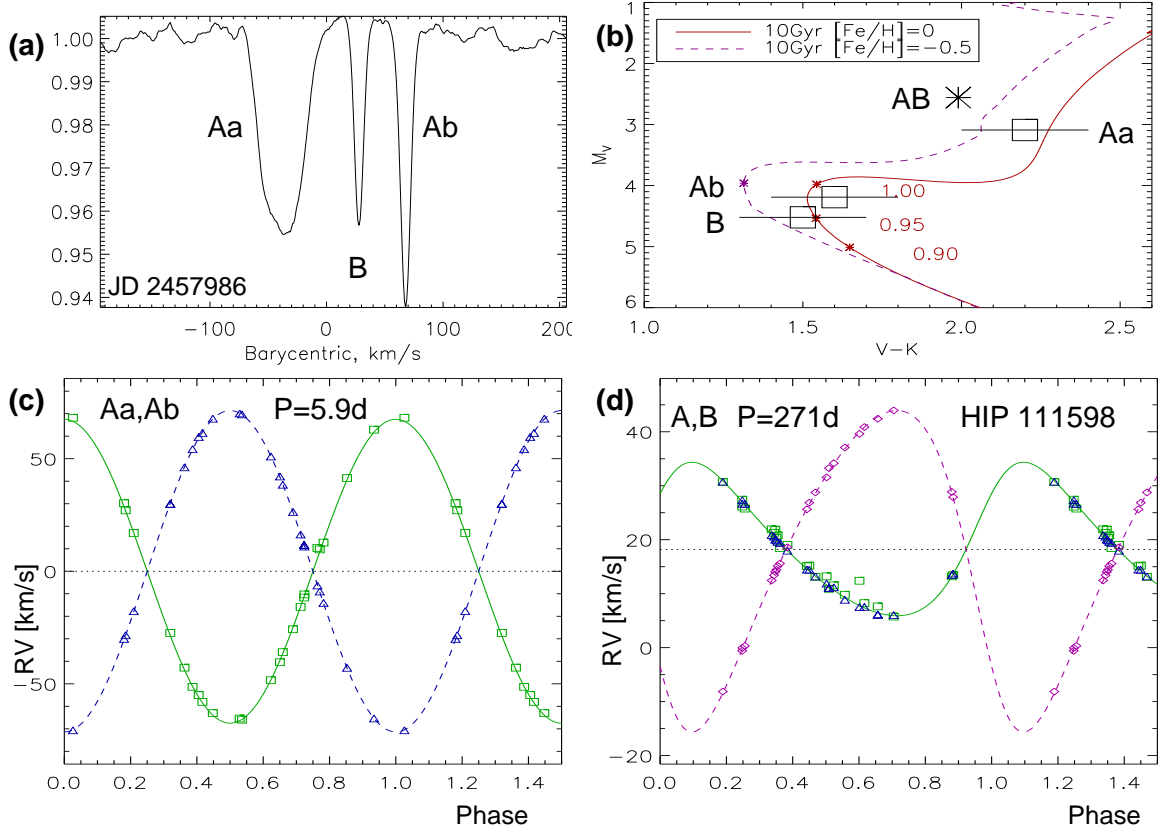}
\caption{The triple system  HIP 111598. (a)  A typical CCF with a triple dip
  recorded on  JD 2457986. (b) Approximate location  of the components
  on the  CMD. The 10-Gyr isochrones  from \citet{Dotter2008}
  (asterisks mark the  masses) are plotted (full line
  for solar metallicity, dashed line for [Fe/H]=$-$0.5). The large
  asterisk corresponds to the combined light of AB.  (c)  The RV
  curve of the inner pair  Aa,Ab, with the outer orbit subtracted. (d)
  The  RV  curve   of  the  outer  pair  A,B   with  the  inner  orbit
  subtracted. Squares, triangles, and diamonds refer to Aa, Ab, and B,
  respectively. 
\label{fig:111598}
}
\end{figure*}

Triple  lines  in   the  spectrum  of  this  star   were  detected  by
\citet{Guillout2009}  in   the  survey   of  X-ray  sources.    On  JD
2452482.60693, they  measured the  RVs of  43 and 3  \kms, but  do not
indicate to  which components they refer.   High chromospheric activity
and  location well  above  the main  sequence  in the  color-magnitude
diagram  (CMD)  suggested  that  this  system  could  be  young.   The
photometric  period of  5.827  days was  found by  \citet{Kiraga2012};
small amplitudes, 0.039  mag in the $V$ band and 0.025  mag in the $I$
band, indicate that the variability is caused by starspots.

The  CHIRON   spectra  show  triple  lines,  all   with  variable  RVs
(Figure~\ref{fig:111598},  a).  The  weakest narrow  lines, attributed
here to the  component B, move relatively slowly with  a period of 271
days.   The other  two components  Aa and  Ab have  stronger  lines of
unequal width  and move much faster,  with a period of  5.87 days. The
elements of both orbits were fitted jointly using the {\tt orbit3.pro}
IDL           code           \citep{TL2017,orbit3}.\footnote{Codebase:
  \url{https://doi.org/10.5281/zenodo.321854}}    The   weighted   rms
residuals  for  Aa,  Ab,  and   B  are  0.91,  0.23,  and  0.24  \kms,
respectively.  The large residuals of Aa are explained by its wide CCF
profile,    likely  distorted by starspots. Both orbits
are plotted in Figure~\ref{fig:111598} (c,d). With the period ratio of
46,  the  interaction  between  the  inner  and  outer  pairs  is  not
negligible.  However, our observations  obtained over a limited period
of  time are  well  represented  by two  Keplerian  orbits with  fixed
elements.  Detailed  dynamical modeling of this system  is outside the
scope of this  paper.  The minimum component masses $  M \sin^3 i$ are
0.835, 0.790 \msun  for the inner pair and 1.492,  0.712 for the outer
orbit.  The mass ratios of  0.947 and 0.476 define the relative masses
of Aa, Ab, and B as 1:0.95:0.93.

A star of one solar radius rotating synchronously with the inner orbit
would have  an equatorial  speed of 8.55  \kms, whereas  the projected
rotation of Aa indicates $V \sin i \sim 27$ \kms. Hence, the radius of
Aa  is   3.2  $R_\odot$  or   larger.  This  component   is  obviously
evolved. The {\it Gaia} DR2 assigns a radius of 3.24 $R_\odot$ and the
effective  temperature of 5440\,K,  treating this  object as  a single
star.  The standard relations for  main sequence stars do not apply to
this  system and,  consequently, the  masses and  temperatures  of the
components can be only guessed. Here, $T_{\rm eff}$ of 5080, 5900, and 5900
K are assumed corresponding to spectral types K0, G0, and G0.  

The combined $V$ magnitude and  the relative fluxes estimated from the
areas of the  CCF dips (with a small correction  for the dependence of
the  CCF  area on  the  assumed  effective  temperature) lead  to  the
relative fluxes of  Aa, Ab, and B  of of 0.61, 0.23, and  0.16 and the
individual  $V$ magnitudes  8.92, 9.97,  and 10.35  mag, respectively.
The  {\it  Gaia}  DR2  parallax of  6.851$\pm$0.063\,mas  defines  the
absolute  magnitudes, assuming zero  extinction. The  individual $V-K$
colors of the components are  not measured directly; I assume here the
$V-K$ colors  of 2.2,  1.6, and  1.5 mag that  match the  combined $K$
magnitude; however, this  is just a guess constrained  by the combined
$V-K$ color.   Figure~\ref{fig:111598} (b)  shows the location  of the
components  on the CMD  (the absolute  magnitudes $M_V$  are measured,
while the  $V-K$ colors are  guessed).  For reference,  two isochrones
from  \citet{Dotter2008} for  the age  of  10\,Gyr and  two values  of
[Fe/H] are plotted. 

 Considering  the isochrones,   the components of  HIP~111598 are
old and, likely, moderately  metal-poor stars (see below), while
their masses  are close  to 1  \msun. They are  located near  the main
sequence turn-off: Aa  has expanded while Ab and B  are still close to
the main sequence.  No attempt of a quantitative match with isochrones
is  made here  because  the knowledge  of  metallicity and  individual
colors (or  temperatures) is still lacking (the  turn-off mass depends
on the  age and metallicity).   The luminosities  and assumed effective
temperatures  correspond to  the stellar  radii of  3.2, 1.3,  and 1.1
$R_\odot$.

\begin{figure}
\epsscale{1.0}
\plotone{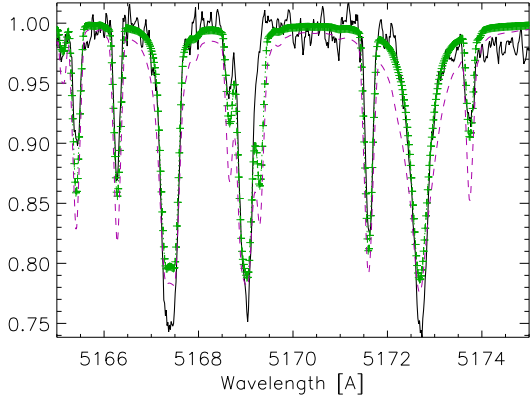}
\caption{Fragment of the disentangled  spectrum of HIP 111598 Ab (full
  line) is  compared to  the  synthetic spectra  with $T_e  = 5900$\,K,
  $\log  g =  3.5$, [Fe/H]=$-$0.5  dex  (green line  and crosses)  and
  [Fe/H]=0 (magenta  dash)  in  the region around  Mg Ib  triplet.  The
  synthetic spectra are scaled for the relative flux of 0.25 to account
  for the dilution by other  components.  Both spectra are smoothed by
  a 4  \kms boxcar.  The lack  of pressure broadening  in strong lines
  suggests a low gravity and a sub-solar metallicity.
\label{fig:syn}
}
\end{figure}

In principle,  the spectra of individual components  can be determined
and modeled using spectral disentangling \citep{Hadrava2009}. However,
this procedure is  not straightforward, computationally intensive, and
designed for binaries rather than  triples; moreover, the author is not
an expert in  this field. To isolate the spectrum  of one component, I
use  the  normalized  spectra  with  merged echelle  orders  and  take
advantage  of  the  measured  RVs.   The  spectra  are  co-added  with
centering  on the  selected component,  while the  moving  features of
other  components are  washed out  by averaging.   Using  these initial
co-added spectra  of the  components as a  first guess, the  result is
refined  iteratively by  subtracting contributions  of  all components
except one,  shifting and averaging again.   The disentangled spectrum
of the  sharp-lined component  Ab is compared  in Figure~\ref{fig:syn}
with  the  synthetic  spectra form  \citet{Palacios2010}\footnote{See
  \url{http://pollux.oreme.org}.}   with the effective  temperature of
5900~K, $\log  g = 3.5$,    solar and  sub-solar metallicity. The
agreement is  not very good, with  some lines being  stronger and some
other  weaker. Synthetic spectra  with $\log  g =  4.5$ or  with solar
metallicity  have  wider  lines  and do  not  match the  observed
spectrum of Ab.

If  the mass sum  of Aa  and Ab  is $\sim$2  \msun, the  inner orbital
radius is  17 $R_\odot$ and  the Roche radius  of the component  Aa is
about 7 $R_\odot$,  so this binary is still  detached.  The absence of
eclipses  constrains  the  inner   inclination  to  $i_{\rm  Aa,Ab}  <
78\fdg5$.   The  orbital  inclinations  cannot  be  evaluated  without
reliable  mass  estimates.   However,  the  similarity  of  the  inner
spectroscopic mass  sum, $M_{\rm Aa+Ab}  \sin^3 i_{\rm Aa,Ab}  = 1.62$
\msun, and the outer primary  mass $M_{\rm Aa+Ab} \sin^3 i_{\rm A,B} =
1.49$   \msun,  suggests   that  the   orbital  inclinations   can  be
similar. Therefore, the inner and outer orbits could be quasi-coplanar.

This  compact triple  system is  rather unusual  in  several respects.
First,  one expects  the components  in the  close 5.9  day  binary to
rotate synchronously with the orbit.  The large star Aa indeed rotates
almost  synchronously,   as  evidenced  by   the  photometric  period.
However, the less evolved star Ab has narrow lines corresponding to $V
\sin i \approx 4.5$ \kms,  whereas the synchronous equatorial speed of
a star with $R = 1.3 R_\odot$ is 11 \kms.  Either Ab indeed rotates slowly or
its axis has a small inclination, hence is misaligned with the orbit. In
both cases,  rapid tidal alignment is expected. Therefore, this
triple system has acquired its present-day architecture only recently.
Another unusual characteristic of this  system is its rather short outer
period of 271 d. Such compact triple systems are rare, and they are
predominantly  coplanar \citep{Borkovits2016}. 

This system   appears to be  relatively old, as  suggested by the
large radius  of the star Aa  and the absence of  the 6707\AA ~lithium
line.  Its spatial velocity  $(U,V,W) = (10.7, -12.4, -30.8)$ \kms
  is typical  for Galactic disk  population.  The  high chromospheric
activity of Aa is likely caused by its synchronization with the orbit.
The slow rotation  of B, $V \sin  i = 1.9$ \kms, also  suggests an old
age.  However, in an old close  binary the component Ab should also be
synchronized with the orbit, which is not the case here.

The  semimajor axis of  the outer  pair A,B  is 6\,mas,  so it  can be
resolved  with long-baseline  interferometers. Accurate  astrometry of
the  outer pair can  reveal the  orientation of  the inner  orbit (its
semimajor axis  is 0.6 mas),  defining the true mutual  inclination in
this interesting system.  {\it Gaia} should detect photocentric motion
of the star, as its estimated amplitude is about 2.5 mas. In fact, the
goodness-of-fit  parameter {\tt  gofAL} in  {\it Gaia}  DR2  is large,
10.08,  indicating  the detection  of  the  astrometric signal.   
  \citet{Brandt2018}  derived  a  highly significant  acceleration  of
  $(-0.6,+1.4)$ mas~yr$^{-1}$.

\section{Summary}
\label{sec:sum}

\begin{figure}
\epsscale{1.0}
\plotone{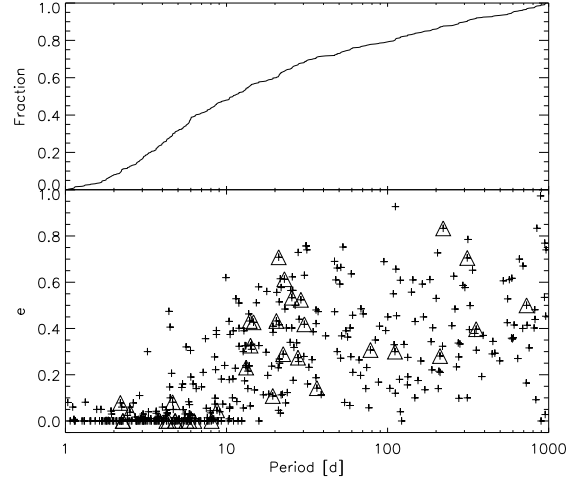}
\caption{Period-eccentricity   relation   for   inner  subsystems   in
  solar-type hierarchies from the MSC. Orbits from papers 1--6 of this
  series are  marked by  triangles. The upper  plot is  the cumulative
  distribution of periods.
\label{fig:pe}
}
\end{figure}

Historically,  close  subsystems in  visual  binaries were  discovered
photometrically    (by    eclipses)    and   then    often    followed
spectroscopically.   This  produced   a  strong  bias  favoring  short
periods. Figure~1 of \citet{TS02} shows  a sharp drop in the number of
inner subsystems with $P > 7$\,d known at the time. However, their own
work  on  spectroscopic  orbits  resulted  in a  more  uniform  period
distribution. The maximum at $P<7$ d was interpreted as a signature of
Kozai-Lidov  cycles with  tides,  in agreement  with the  distribution
expected         from         the         population         synthesis
\citep{Fabrycky2007,Hamers2019}.

Nowadays  the situation  has changed  dramatically because  most known
subsystems in solar-type hierarchies were discovered spectroscopically
(by variable  RVs or double  lines) rather than  photometrically. This
method works in  a much wider range of  inner periods. Unknown orbital
elements of  spectroscopic subsystems  discovered from a  few spectra,
e.g. by  \citet{N04}, are determined by the  follow-up work, including
this series of papers.

Figure~\ref{fig:pe} shows  the period-eccentricity plot  for 528 inner
subsystem with  primary mass  less than 1.5  \msun extracted  from the
MSC. The  36 orbits resulting  from this series  (a modest 7\%  of the
total) are  highlighted by the triangles.  The cumulative distribution
of periods  is also  plotted.  Quite remarkably,  this curve  shows no
details  suggesting  a  local maximum  at  $P \sim  10$  d.   Instead,  this
distribution is smooth.  Of course,  the sample extracted from the MSC
is   biased, being  derived from heterogeneous  sources. However,
the lack of the tidal signature in the period distribution is obvious.

The preference of  close binaries to be inner  members of hierarchical
systems  is  a  well  established  fact. Overall,  the  statistics  of
hierarchical  multiplicity in  the  67-pc sample  of solar-type  stars
matches independent  combination of  inner and outer  periods selected
from  the same distribution  and filtered  by the  dynamical stability
criterion.   However, this  model does  not work  for  subsystems with
$P<10$  d because their  number is  larger than  would result  from an
independent  selection  \citep[see  Fig.~11 in][]{FG67b}.   Therefore,
formation of  close binaries  within hierarchies is  somehow enhanced,
but is not necessarily related to tides.

An  alternative  channel  of  close  binary formation  is  the  inward
migration  of wider  pairs driven  by  accretion.  In  this case,  the
presence of outer  companion(s) is, on one hand,  a consequence of the
accretion  (availability  of  the   mass  supply  sufficient  to  form
additional companions)  and, on  the other hand,  a factor  that might
help migration by supplying mis-aligned  gas to the inner binary. Mass
ratios  of   accreting  binaries   always  increase  and   often  tend
asymptotically  toward  one (twins).  Indeed,  many inner  subsystems
studied here are twins.

\acknowledgements

I thank the operator of  the 1.5-m telescope R.~Hinohosa for executing
observations  of  this  program  and  L.~Paredes  for  scheduling  and
pipeline processing.  Comments by  the Referee, K.~Fuhrmann, helped to
improve the manuscript.

This work  used the  SIMBAD service operated  by Centre  des Donn\'ees
Stellaires  (Strasbourg, France),  bibliographic  references from  the
Astrophysics Data  System maintained  by SAO/NASA, and  the Washington
Double Star Catalog maintained at USNO.
This work has made use of data from the European Space Agency (ESA) mission
{\it Gaia} (\url{https://www.cosmos.esa.int/gaia}), processed by the {\it Gaia}
Data Processing and Analysis Consortium (DPAC,
\url{https://www.cosmos.esa.int/web/gaia/dpac/consortium}). Funding for the DPAC
has been provided by national institutions, in particular the institutions
participating in the {\it Gaia} Multilateral Agreement.

{\it Facility:}  \facility{CTIO:1.5m}











\begin{thebibliography}{99}

\bibitem [Borkovits et al.(2016)]  {Borkovits2016}
Borkovits, T., Hajdu, T., Sztakovics, J,,  et al. 2016, MNRAS, 455, 4136

\bibitem[Boro Saikia et al.(2018)]{Boro2018}
Boro Saikia, S., Marvin, C. J., Jeffers, S. V. et al. 2018, A\&A, 616, 108

\bibitem[Brandt(2018)]{Brandt2018}
Brandt, T. D. 2018, ApJS, 239, 31


\bibitem[Desidera et al.(2006)]{Desidera2006}
Desidera, S., Gratton, R. G., Lucatello, S. et al. 2006, A\&A, 454, 553


\bibitem[Dotter et al.(2008)]{Dotter2008} 
         Dotter, A., Chaboyer, B., Jevremovi\'c, D. et al. 2008,
         \apjs, 178, 89

\bibitem[Eker et al.(2008)]{Eker2008}
Eker, Z., Filiz Ak, N., Bilir, S. et al. 2008, MNRAS, 389, 1722

\bibitem[Evans(1968)]{Evans1968}
Evans, D. S. 1968, QJRAS, 9, 388

\bibitem[Fabrycky  \& Tremaine (2007)]{Fabrycky2007}
Fabrycky D. \& Tremaine S., 2007, ApJ, 669, 1298

\bibitem[Fekel(1981)]{Fekel1981}
Fekel, F. C. 1981, ApJ, 246, 879

\bibitem[Fuhrmann \& Chini(2019)]{Fuhrman2019}
Fuhrmann, K. \& Chini, R. 2019, MNRAS, 482, 471


\bibitem [Gaia collaboration(2018)]  {Gaia}
Gaia Collaboration, Brown, A. G. A., Vallenari, A., Prusti, T. et
al. 2018, A\&A, 595, 2 (Vizier Catalog 	I/345/gaia2).

\bibitem[Gomez et al.(2016)]{Doc2016a}
Gomez, J., Docobo, J. A., Campo, P. P. \& Mendez, R. A. 2016, AJ, 152, 216


\bibitem[Guillout et al.(2009)]{Guillout2009}
Guillout, P.,  Klutsch, A.,  Frasca, A. et al. 2009, A\&A, 504, 829

\bibitem[Hadrava(2009)]{Hadrava2009}
Hadrava, P. 2009, A\&A, 494, 399

\bibitem[Hamers(2019)]{Hamers2019}
Hamers, A. 2019, MNRAS, 482, 2262

\bibitem[Jenkins et al.(2015)]{Jenkins2015}
Jenkins, J. S., D\'{i}az, M., Jones, H. R. A.  et al. 2015, MNRAS, 453, 1439


\bibitem [Kiraga(2012)]  {Kiraga2012}
Kiraga M., 2012, Acta Astron., 62, 67


\bibitem[Mason et al.(2001)]{WDS}
Mason, B. D., Wycoff, G. L., Hartkopf, W. I., Douglass, G. G. \&
Worley, C. E. 2001, AJ, 122, 3466 (WDS)

\bibitem[Moe \& Kratter(2018)]{Moe2018}
Moe, M. \& Kratter, K. M. 2018, ApJ, 854, 44


\bibitem[Nordstrom \& Andersen(1985)]{N85}
Nordstrom, B. \& Andersen, J. 1985, A\&AS, 61, 53

\bibitem[Nordstr\"om et al.(2004)]{N04}
Nordstr\"om,  B., Mayor,  M.,  Andersen, J.  et al.  2004, A\&A, 418, 989 


\bibitem[Palacios et al.(2010)]{Palacios2010}
 Palacios A. , Gebran M., Josselin E. et. al 
2010, A\&A, 516, 13

\bibitem[Plavchan et al.(2009)]{Plavchan2009}
Plavchan, P., Werner, M. W., Chen, C. H. et al. 2009, ApJ, 698, 1068 

\bibitem[Pojmanski(1997)]{ASAS}
Pojmanski, G., 1997, Acta Astronomica, 47, 467.


\bibitem[Tokovinin(2014)]{FG67b}
 Tokovinin, A.  AJ, 2014, 147, 87 

\bibitem[Tokovinin(2015)]{survey}
Tokovinin, A.  2015, AJ,   150, 177

\bibitem[Tokovinin(2016a)]{paper1}
Tokovinin, A.  2016a, AJ, 152, 11 (Paper 1)

\bibitem[Tokovinin(2016b)]{paper2}
Tokovinin, A.  2016b, AJ, 152, 10 (Paper 2)

\bibitem[Tokovinin(2016c)]{orbit}
Tokovinin, A. 2016c, ORBIT: IDL Software for Visual, Spectroscopic, and
Combined Orbits, Zenodo, \url{doi:10.2581/zenodo.61119}

\bibitem[Tokovinin(2017)]{orbit3}
Tokovinin, A. 2017, ORBIT3—Orbits of Triple Stars, Zenodo, 
\url{doi:10.5281/zenodo.321854}

\bibitem[Tokovinin(2018a)]{MSC}
Tokovinin, A. 2018a, ApJS,  235, 6

\bibitem[Tokovinin(2018b)]{paper3}
Tokovinin, A. 2018b, AJ, 156, 48 (Paper 3)

\bibitem[Tokovinin(2018c)]{paper4}
Tokovinin, A. 2018c, AJ, 156, 194 (Paper 4)

\bibitem[Tokovinin(2019)]{paper5}
Tokovinin, A. 2019, AJ, 157, 91 (Paper 5)

\bibitem[Tokovinin et al.(2018)]{young}
Tokovinin, A., Corbett, H., Fors, O., et al. 2018, AJ, 156, 120

\bibitem[Tokovinin et al.(2013)]{CHIRON}
Tokovinin, A., Fischer, D. A., Bonati, M. et al. 2013, PASP, 125, 1336

\bibitem[Tokovinin \& Latham(2017)]{TL2017}
Tokovinin, A. \& Latham, D. W.  2017, ApJ, 838, 54

\bibitem[Tokovinin et al.(2019)]{SOAR}
Tokovinin, A., Mason, B. D., Mendez, R. A. et al. 2019,  AJ, 158, 48

\bibitem[Tokovinin et al.(2015)]{LCO}
Tokovinin, A., Pribulla, T., \& Fischer, D.  2015, AJ, 149, 8

\bibitem[Tokovinin \& Smekhov(2002)]{TS02}
Tokovinin, A. A. \& Smekhov, M. G. 2002, A\&A, 382,  118

\bibitem[van Leeuwen(2007)] {HIP2}
van Leeuwen, F. 2007, A\&A, 474, 653 (HIP2)





\end{thebibliography}
\end{document}